\documentclass{article}

\usepackage[top=2.75cm, bottom=2.0cm, left=3.00cm, right=2.50cm]{geometry}
\usepackage[hyphens]{url}
\usepackage{amsmath}
\usepackage{amsthm}
\usepackage{amsfonts}
\usepackage{float}
\usepackage{colortbl}
\usepackage{framed}
\usepackage{enumitem}
\usepackage{tikz}
\usepackage{multirow}
\usepackage{appendix}
\usepackage{enumitem}
\usepackage{caption}
\usepackage{algorithm}
\usepackage{hyperref}
\usepackage{tikz}
\usepackage{mathtools}
\usepackage{amsmath}
\usepackage{amsfonts}
\usepackage{graphicx}
\usepackage{subcaption}
\usepackage{times}
\usepackage{algorithm}
\usepackage{algpseudocode}
\usepackage{extarrows}
\usepackage{turnstile}
\usepackage{wrapfig}

\hypersetup{
    colorlinks=true,
    linkcolor=blue,
    filecolor=magenta,      
    urlcolor=cyan,
}

\providecommand{\citep}{\cite}
\newcommand{\comment}[1]{// #1}

\newtheorem{definition}{Definition}
\newtheorem{theorem}{Theorem}

\newcommand{\share}[2]{{#1}^{(#2)}}  
\newcommand{\vc}[2]{{#1}_{#2}}  
\newcommand{\vcs}[3]{\share{#1}{#3}_{#2}}  
\newcommand{\si}[2]{#1 [\![ #2 ]\!] }  
\newcommand{\so}[2]{#1 [\![ #2 ]\!] }  

\newcommand{\genT}{\ensuremath{\mathcal{G}}}  
\newcommand{\voterT}{\ensuremath{\mathcal{V}}}  
\newcommand{\poT}{\ensuremath{\mathcal{P}}}  
\newcommand{\devT}{\ensuremath{\mathcal{D}}}  
\newcommand{\bboxT}{\ensuremath{\mathcal{B}}}  
\newcommand{\mixerT}{\ensuremath{\mathcal{M}}}  
\newcommand{\gen}[1]{\ensuremath{\genT_{#1}}}  
\newcommand{\voter}[1]{\ensuremath{\voterT_{#1}}}  
\newcommand{\po}[1]{\ensuremath{\poT_{#1}}}  
\newcommand{\dev}[1]{\ensuremath{\devT_{#1}}}  
\newcommand{\bbox}[1]{\ensuremath{\bboxT_{#1}}}  
\newcommand{\mixer}[1]{\ensuremath{\mixerT_{#1}}}  
\newcommand{\igen}{\ensuremath{t}}  
\newcommand{\ngens}{\ensuremath{\tau}}  
\newcommand{\ivoter}{\ensuremath{i}}  
\newcommand{\nvoters}{\ensuremath{n}}  
\newcommand{\ibooth}{\ensuremath{j}}  
\newcommand{\nbooths}{\ensuremath{\ell}}  
\newcommand{\imixer}{\ensuremath{k}}  
\newcommand{\nmixers}{\ensuremath{m}}  
\newcommand{\querier}{\ensuremath{Q}}  
\newcommand{\gens}{\ensuremath{(\gen{\igen})_{\igen \in [\ngens]}}}  
\newcommand{\voters}{\ensuremath{(\voter{\ivoter})_{\ivoter \in [\nvoters]}}}  
\newcommand{\pos}{\ensuremath{(\po{\ibooth})_{\ibooth \in [\nbooths]}}}  
\newcommand{\devs}{\ensuremath{(\dev{\ibooth})_{\ibooth \in [\nbooths]}}}  
\newcommand{\bboxes}{\ensuremath{(\bbox{\ibooth})_{\ibooth \in [\nbooths]}}}  
\newcommand{\mixers}{\ensuremath{(\mixer{\imixer})_{\imixer \in [\nmixers]}}}  
\newcommand{\auditor}{\ensuremath{\mathcal{A}}}  

\newcommand{\setup}{\ensuremath{\mathsf{Setup}}}  
\newcommand{\ballotgen}{\ensuremath{\mathsf{BallotGen}}}  
\newcommand{\cast}{\ensuremath{\mathsf{Cast}}}  
\newcommand{\tally}{\ensuremath{\mathsf{Tally}}}  
\newcommand{\ballotaudit}{\ensuremath{\mathsf{BallotAudit}}}  
\newcommand{\receiptaudit}{\ensuremath{\mathsf{ReceiptAudit}}}  
\newcommand{\audit}{\ensuremath{\mathsf{TallyAudit}}}  

\newcommand{\sign}{\ensuremath{\mathsf{Sign}}}
\newcommand{\verify}{\ensuremath{\mathsf{Ver}}}
\newcommand{\dec}{\ensuremath{\mathsf{Dec}}}
\newcommand{\sharealg}{\ensuremath{\mathsf{Share}}}  

\newcommand{\bbr}{\ensuremath{R}}  
\newcommand{\bbc}{\ensuremath{C}}  
\newcommand{\bbv}{\ensuremath{V}}  
\newcommand{\bbp}{\ensuremath{P}}  
\newcommand{\vid}{\ensuremath{\mathsf{vid}}}  
\newcommand{\bid}{\ensuremath{\mathsf{bid}}}  
\newcommand{\booth}{\ensuremath{\mathsf{booth}}}  
\newcommand{\pkey}[1]{\ensuremath{\mathsf{pk}_{#1}}}  
\newcommand{\skey}[1]{\ensuremath{\mathsf{sk}_{#1}}}  
\newcommand{\ballot}[2]{\ensuremath{\vc{b}{(#1, #2)}}}  
\newcommand{\iballot}{\ensuremath{\beta}}  
\newcommand{\nballots}[1]{\ensuremath{n_{\mathsf{b}_{#1}}}}  
\newcommand{\ballotsbooth}[1]{\ensuremath{(\ballot{\iballot}{#1})_{\iballot \in [\nballots{#1}]}}}  
\newcommand{\ballots}[1]{\ensuremath{(\ballotsbooth{\ibooth})_{\ibooth \in [\nbooths]}}}  

\newcommand{\pkbody}[2]{\ensuremath{\{(#1): #2\}}}  
\newcommand{\nizkpktxt}{\ensuremath{\mathsf{NIZKPK}}}
\newcommand{\nizkpk}[2]{\ensuremath{\nizkpktxt\pkbody{#1}{#2}}}  
\newcommand{\tm}{\ensuremath{\Pi_{\mathsf{TM}}}}  
\newcommand{\pks}{\ensuremath{\Pi_{\mathsf{S}}}}  

\newcommand{\Zq}{\mathbb{Z}_{q}}  
\newcommand{\Gone}{\mathbb{G}_{1}}  
\newcommand{\Gtwo}{\mathbb{G}_{2}}  
\newcommand{\GT}{\mathbb{G}_{T}}  

\newcommand{\keygen}{\ensuremath{\mathsf{Keygen}}}  
\newcommand{\enc}{\ensuremath{\mathsf{Enc}}}  
\newcommand{\mix}{\ensuremath{\mathsf{Mix}}}  
\newcommand{\btrin}{\ensuremath{\mathsf{BTraceIn}}}  
\newcommand{\btrout}{\ensuremath{\mathsf{BTraceOut}}}  
\newcommand{\encscheme}{\ensuremath{\mathsf{E}}}
\newcommand{\renc}{\ensuremath{\mathsf{REnc}}}
\newcommand{\wit}[1]{\share{\omega}{#1}}                  

\newcommand{\pke}{\ensuremath{\encscheme}}

\newcommand{\pkee}{\ensuremath{\pke.\enc}}

\newcommand{\pked}{\ensuremath{\pke.\dec}}

\newcommand{\mpk}{\ensuremath{\mathsf{mpk}}}
\newcommand{\msk}[1]{\ensuremath{\share{\mathsf{msk}}{#1}}}
\newcommand{\theztxt}{\ensuremath{\mathsf{Pa}}}
\newcommand{\thez}{\ensuremath{\encscheme_{\theztxt}^{\mathsf{th}}}}

\newcommand{\theze}{\ensuremath{\thez.\enc}}
\newcommand{\thezre}{\ensuremath{\thez.\renc}}

\newcommand{\thezpk}{\ensuremath{\mathsf{pk}_{\theztxt}}}

\newcommand{\thegtxt}{\ensuremath{\mathsf{EG}}}
\newcommand{\theg}{\ensuremath{\encscheme_{\thegtxt}^{\mathsf{th}}}}

\newcommand{\thegpk}{\ensuremath{\mathsf{pk}_{\thegtxt}}}

\newcommand{\pokcomm}{\ensuremath{\rho_{\gamma}}}

\newcommand{\FO}{\ensuremath{{\mathsf{FO}}}} 

\newcommand{\goodbooths}{\ensuremath{{\mathsf{goodbooths}}}}
\newcommand{\badbooths}{\ensuremath{{\mathsf{badbooths}}}}

\newcommand{\iindices}[1]{\ensuremath{I_{#1}}}
\newcommand{\jindices}[1]{\ensuremath{I'_{#1}}}

\newcommand{\R}{\ensuremath{{R}}}
\newcommand{\C}{\ensuremath{{C}}}
\newcommand{\V}{\ensuremath{{V}}}
\newcommand{\Pa}{\ensuremath{{P}}}

\title{\emph{OpenVoting}: Recoverability from Failures in Dual Voting}

\author{
  Prashant Agrawal\thanks{Department of Computer Science and Engineering, IIT Delhi, New Delhi, India. Email: {\it \{prashant,svs,suban\}@cse.iitd.ac.in}}
  \and
  Kabir Tomer\thanks{Department of Computer Science, University of Illinois Urbana-Champaign. Email: {\it ktomer2@illinois.edu.in}. Work done when at IIT Delhi.}
  \and
   Abhinav Nakarmi\thanks{Department of Computer Science and Centre for Digitalisation, AI and Society, Ashoka University, Sonipat, India.  Email: {\it abhinav.nakarmi@alumni.ashoka.edu.in, \{mahavir.jhawar,subhashis.banerjee\}@ashoka.edu.in}}
   \and
   Mahabir Prasad Jhanwar\footnotemark[3]
   \and
   Subodh Sharma\footnotemark[1]
   \and
   Subhashis Banerjee\footnotemark[1] \footnotemark[3]
 }
\usepackage{algpseudocode}
\usepackage{algorithm}

\begin{document}

\maketitle


\begin{abstract}
	In this paper we address the problem of recovery from failures  without re-running entire elections when elections fail to verify. We consider the setting of \emph{dual voting} protocols, where the cryptographic guarantees of end-to-end verifiable voting (E2E-V) are combined with the simplicity of audit using voter-verified paper records (VVPR). We first consider the design requirements of such a system and then suggest a protocol called \emph{OpenVoting}, which identifies a verifiable subset of error-free votes consistent with the VVPRs, and the polling booths corresponding to the votes that fail to verify with possible reasons for the failures. To an ordinary voter \emph{OpenVoting} looks just like an old fashioned paper based voting system, with minimal additional cognitive overload.
\end{abstract}

\section{Introduction}

Conducting large-scale public elections in a dispute-free manner is not an easy task. On the one hand, there are end-to-end verifiable voting  (E2E-V) systems \citep{scratchandvote,punchscan,pretavoter,scantigrity,starvote} that provide cryptographic guarantees of correctness. Although the guarantees are sound, these systems are not yet very popular in large public elections. As the German Constitutional Court  observes \citep{germany}, depending solely on cryptographic guarantees is somewhat untenable because verification of  election results requires expert knowledge. Moreover, in case voter checks or universal verifications fail, the E2E-V systems do not provide easy methods of recovery without necessitating complete re-election \citep{Bernhardsurvey}.

On the other hand, there are systems that rely on paper-audit trails to verify electronic tallies \citep{stark-conservative-statistical,stark-super-simple,bravo}. These systems maintain reliable records of cleartext voter-marked paper ballots or voter-verified paper records (VVPRs) in addition to electronic vote records. They use electronic counting for efficiency and conduct easy-to-understand statistical audits, called {\em risk-limiting audits} (RLAs), to demonstrate that the electronic winners match the winners that would be declared by a full paper count. In case of conflict, the electronic outcome is suggested to be replaced by the paper one. However, these systems require the electorate to trust that the paper records correctly represent voter intent and are not corrupted in the custody chain from the time of voting to that of counting or auditing.

\emph{Dual voting} approaches, where the voting protocols support simultaneous voting for both the cryptographic and the VVPR-based systems \citep{Benaloh2008,wombat,sigma-ballots,pretavoter-hrpat,starvote,single-layer-osv}, combine the cryptographic guarantees of E2E-V systems with the simplicity and adoptability of paper records. However, in most existing dual voting systems, one typically ends up running two parallel and independent elections, only coupled loosely through simultaneous voting for both in the polling booth. If the electronic and paper record systems are not tightly coupled, and demonstrably in one-to-one correspondence, then it begs the questions: which ought be the legal definition of the vote, and, in case of a tally mismatch, which should be trusted? Why? And how to recover from errors?

It appears that existing approaches either do not provide any recovery mechanism or recover by privileging VVPR counts over electronic counts. In large public elections running simultaneously at multiple polling booths per constituency, failures due to intended or unintended errors by different actors are expected. Polling officers may upload wrong encrypted votes, backend servers may decrypt votes incorrectly, paper records may be tampered with during the custody chain, and voters may put bogus votes in ballot boxes to discredit the election. Discarding the entire election due to failures caused by some bad actors or completely trusting the VVPRs are both unsatisfactory solutions.

In this paper, we study the problem of \emph{recoverability} of a dual voting protocol from audit failures. We consider large, multi-polling booth, first-past-the-post elections like the national elections in India. We observe that except for backend failures, most of the other failures are due to localised corruption of individual polling booths. Therefore, we propose to identify the offending polling booths and perform a local re-election --- if at all required --- only at those polling booths. Errors --- despite the best efforts to minimise them --- are inevitable in large elections and such localised recovery  may considerably improve the election's overall robustness and transparency. 

However, recoverability has a natural tradeoff with vote secrecy. For example, a naive  approach that simply publishes and audits votes for each polling booth reveals voting statistics of each booth. In electoral contexts where voters are assigned a specific polling booth according to their residential neighbourhoods, with  only a few thousand voters per booth, e.g., in India, revealing booth-level voting statistics poses a significant risk of localised targeting and coercion \citep{totaliser}. Our approach minimises booth-level voting data exposure, disclosing only what is absolutely necessary for recovery. \\

\noindent \textbf{Main contributions.} \textbf{1)} We analyse the design requirements for a recoverable and secrecy-preserving dual voting protocol (Section \ref{sec:requirements}). \textbf{2)} We formalise the notion of recoverability and secrecy in terms of the capability to verifiably identify polling booths contributing to verification failures and extract a verifiable subset of error-free votes in zero-knowledge (Section \ref{formalisation}). \textbf{3)} We propose a novel dual-voting protocol called \emph{OpenVoting} that satisfies our notions of recoverability and secrecy (Section \ref{sec:protocol}). \\

\noindent \textbf{Related work.} 
Dual voting was introduced by Benaloh \citep{Benaloh2008}, following which multiple dual voting protocols emerged  \citep{pretavoter-hrpat,wombat,sigma-ballots,starvote,single-layer-osv}. Bernhard et al. \citep{Bernhardsurvey} gives a comprehensive survey of the tradeoffs and open problems in E2E-V and RLA-based voting.

Rivest \citep{softwareindependence} proposed the notion of \emph{strong software independence} that is similar to our notion of recoverability. It demands that a detected change or error in an election outcome (due to a change or error in the software) can be corrected without re-running the (entire) election. However, ``correcting'' errors without re-running even parts of an election requires a ground truth, which is usually assumed to be the paper audit trail. Instead, we propose partial recoverability via fault localisation, without completely trusting either paper or electronic votes. The notion of \emph{accountability} \citep{accountability} is also related, but it is focused on assigning blame for failures and not on recovering from them.

\section{Design Requirements}
\label{sec:requirements}
 
In a typical dual voting protocol, the vote casting process produces $a)$ a VVPR containing the voter's vote in cleartext and $b)$ a voter receipt containing an encryption of the vote. The encrypted votes are published on a bulletin board, typically by a \emph{polling officer}, and are processed by a cryptographic \emph{backend} to produce the electronic tally. The backend typically consists of multiple independent servers which jointly compute the tally from the  encrypted inputs, provide a proof of correctness, and preserve  vote secrecy unless a threshold number of servers are corrupted. VVPRs counted together produce the paper tally.

Our high-level goal is to publicly verify whether both tallies represent true voter intents and whether all public outputs are consistent with each other. If not, the aim of recovery is to identify booths contributing  to the inconsistencies, and segregate the outputs produced by other error-free booths, without leaking any additional information. For this, the protocol design must fundamentally have the following features:
\begin{enumerate}[leftmargin=*]
	\item The backend must publish individual decrypted votes with matching identifiers with the VVPRs\footnote{Homomorphic tallying based backends \citep{scratchandvote,starvote} report only the final tally and do not support this.}, to narrow down tally inconsistencies to individual vote mismatches.
	\item The encrypted votes must have voter and booth identifiers. The former enable matching with voter receipts; the latter enable identifying booths in case of errors.
	\item The decrypted votes and VVPRs and their identifiers must be unlinkable to encrypted votes, voter receipts or voter identifiers to ensure vote secrecy. They should also be unlinkable to the booth identifiers to hide booth-level voting statistics.
	\item For the same reason, VVPRs should be revealed and counted only after aggregating them over all the polling booths.
\end{enumerate}
The encrypted and decrypted votes must be published on two public bulletin boards to enable voters to match their receipts and public verification of the electronic tally. It will also be helpful to upload all VVPRs after scanning, and as many voter receipts as possible, to two other bulletin boards for better transparency and public verifiability. We depict such a design in Figure \ref{fig:artifacts}.

\begin{figure}[t]
	\centering
	\scalebox{0.8}{
		\begin{tabular}{c@{\hskip 0.2cm} c@{\hskip 0.2cm} c@{\hskip 0.2cm} c@{\hskip 0.2cm} c}
			\begin{tabular}{l}
				\hline
				\multicolumn{1}{|c|}{\cellcolor{gray!20} Receipts ($\bbr$)} \\ 
				\hline
				\\ 
				\multicolumn{1}{c}{$\vdots$} \\
				\hline
				\multicolumn{1}{|c|}{$\vid_i, \booth_i, c_i$} \\
				\hline
				\multicolumn{1}{c}{$\vdots$} \\
				\\
			\end{tabular} & 
			\begin{tabular}{|c|}
				\hline
				\cellcolor{gray!20} Encrypted votes ($\bbc$) \\
				\hline
				$\vid_1, \booth_1, c_1$ \\
				\hline
				$\vdots$ \\
				\hline
				$\vid_i, \booth_i, c_i$ \\
				\hline
				$\vdots$ \\
				\hline
				$\vid_{n_{\mathsf{c}}}, \booth_{n_{\mathsf{c}}}, c_{n_{\mathsf{c}}}$ \\
				\hline
			\end{tabular} & 
			\begin{tabular}{c|c|c}
				\multicolumn{1}{c}{} & \multicolumn{1}{c}{$\mixer{1},\dots,\mixer{m}$} & \\
				\cline{2-2}
				$\rightarrow$ &                        & $\rightarrow$ \\
				$\vdots$      & Cryptographic backend  & $\vdots$ \\
				$\rightarrow$ & \multirow{3}{*}{$(\pi)$}                       & $\rightarrow$ \\
				$\vdots$      &                        & $\vdots$ \\
				$\rightarrow$ &                        & $\rightarrow$\\
				\cline{2-2}
			\end{tabular} &
			\begin{tabular}{|c|}
				\hline
				\multicolumn{1}{|c|}{\cellcolor{gray!20} Decrypted votes ($\bbv$)} \\
				\hline
				$\bid'_1, v'_1$ \\
				\hline
				$\vdots$ \\
				\hline
				$\bid'_i, v'_{i}$ \\
				\hline
				$\vdots$ \\
				\hline
				$\bid'_{n_{\mathsf{v}}}, v'_{n_{\mathsf{v}}}$ \\
				\hline
			\end{tabular} &
			\begin{tabular}{l}
				\hline
				\multicolumn{1}{|c|}{\cellcolor{gray!20} VVPRs ($\bbp$)} \\ 
				\hline
				\\
				\multicolumn{1}{c}{$\vdots$} \\
				\hline
				\multicolumn{1}{|c|}{$\bid_{\pi(i)}, v_{\pi(i)}$} \\
				\hline
				\multicolumn{1}{c}{$\vdots$} \\
				\\
			\end{tabular}
		\end{tabular}
	}
	\caption{{\small A recoverable dual voting protocol design. The VVPR for a voter with identifier $\vid_i$ voting at booth $\booth_i$ contains a ballot identifier $\bid_i$ and cleartext vote $v_i$. Her encrypted vote $c_i$ encrypts a value, e.g., $(\bid_i,v_i)$, that when decrypted can be uniquely matched with the corresponding VVPR. Decrypted votes are published by backend servers $\mixer{1}, \dots,\mixer{m}$ in a permuted order under a secret shared permutation $\pi$ such that $(\bid'_{i}, \vc{v}{i}')$ $=$ $(\bid_{\pi(i)}, \vc{v}{\pi(i)})$. Note that $n_\mathsf{c}$ and $n_\mathsf{v}$ denote the number of encrypted votes and decrypted votes respectively.}}
	\label{fig:artifacts}
\end{figure}

\begin{figure}[t]
	\centering
	\scalebox{0.8}{
		\begin{tabular}{l l }
			\hline
			\multicolumn{2}{l}{\textbf{Input-phase failures}\footnotemark[1]$^{,}$\footnotemark[2]} \\
			$\mathsf{FI}_1$ & A receipt $r$ against $\vid$ exists in $\bbr$ but no encrypted vote against $\vid$ exists in $\bbc$ \\
			$\mathsf{FI}_2$ & The encrypted vote $c$ in $\bbc$ against $\vid$ does not match the receipt $r$ in $\bbr$ against $\vid$  \\
			\hline
			\multicolumn{2}{l}{\textbf{Mixing-phase failures}\footnotemark[2]$^{,}$\footnotemark[3]} \\
			$\mathsf{FM}_1$ & An encrypted vote $c$ in $\bbc$ does not decrypt to any cleartext vote $(\bid,v)$ in $\bbv$ \\
			$\mathsf{FM}_2$ & A cleartext vote $(\bid,v)$ in $\bbv$ is not obtained by decrypting any encrypted vote $c$ in $\bbc$ \\
			$\mathsf{FM}_3$ & Two encrypted votes in $\bbc$ decrypt to the same cleartext vote $(\bid,v)$ in $\bbv$ \\
			\hline
			\multicolumn{2}{l}{\textbf{Output-phase failures}\footnotemark[2]} \\
			$\mathsf{FO}_1$ & An (electronic) decrypted vote $(\bid,v)$ exists in $\bbv$ but no VVPR against $\bid$ exists in $\bbp$ \\
			$\mathsf{FO}_2$ & A VVPR $(\bid,v)$ exists in $\bbp$ but no decrypted vote against $\bid$ exists in $\bbv$ \\
			$\mathsf{FO}_3$ & The decrypted vote $v$ against $\bid$ in $\bbv$ does not match the cleartext vote in the VVPR against $\bid$ in $\bbp$ \\
			$\mathsf{FO}_4$ & Two decrypted votes in $\bbv$ match with a single VVPR $(\bid,v)$ in $\bbp$ \\
			$\mathsf{FO}_5$ & Two VVPRs in $\bbp$ match with a single decrypted vote $(\bid,v)$ in $\bbv$ \\
			\hline
			\multicolumn{2}{l}{\textbf{Cast-as-intended failures}} \\
			$\mathsf{FC}$ & A receipt $r$ obtained at a polling booth $j$ does not encrypt the voter's intended vote correctly \\
            \hline
		\end{tabular}
	}
	\begin{flushleft}
		\scriptsize{\textsuperscript{1}A spurious encrypted vote against a $\vid$ in $\bbc$ without a receipt in $\bbr$ against that $\vid$ is not considered a failure, because some voters may not upload their receipts. Also, we do not consider duplicated receipts and encrypted votes because $\vid$s are assumed to be unique identifiers.} \\
		\scriptsize{\textsuperscript{2}We only consider authentic entries in $\bbr$, $\bbc$, $\bbv$ and $\bbp$. Failures where the authenticity of these items cannot be verified are considered equivalent to failures where they are not even uploaded. Receipts and VVPRs are authenticated by official stamps and encrypted and decrypted votes by appropriate digital signatures.} \\
		\scriptsize{\textsuperscript{3}The case of a single encrypted vote in $\bbc$ decrypting to two different entries in $\bbv$ is not considered because this will result in duplicated entries in $\bbv$, which can be clearly attributed to backend failures and removed without any dispute.} \\
	\end{flushleft}
	\caption{{\small Potential failures given public outputs $(\bbr,\bbc,\bbv,\bbp)$.}}
	\label{fig:failures}
\end{figure}

Note that the public outputs in Figure \ref{fig:artifacts} are effectively \emph{claims} endorsed by various entities as to what should be the correct vote: receipts by voters, encrypted votes by polling officers, decrypted votes by the backend servers, and VVPRs by the VVPR counting authorities. We group disputes between these claims into \emph{input-phase failures}, for mismatches between published voter receipts and encrypted votes, \emph{mixing-phase failures}, for mismatches between encrypted votes and decrypted votes, and \emph{output-phase failures}, for mismatches between decrypted votes and VVPRs (see Figure \ref{fig:failures}). Further, we categorise claims of receipts not encrypting voter intents correctly as \emph{cast-as-intended failures}. Given these failures, recoverability requires an audit protocol that verifies whether the different claims for a given vote are consistent, resolves disputes otherwise, and narrows down the affected votes when the disputes are unresolvable. 

To recover from input-phase failures, it is not sufficient if a statistically significant sample of voters from the entire constituency verify their receipts, because in case of any failure, all the uploaded encrypted votes become untrustworthy. Thus, the population for sampling must be each polling booth. This does increase the voter verification overhead, but offers better localisation of errors and recovery.

Recoverability from mixing-phase failures requires that in case the output list of decrypted votes is not correct, individual failing entries --- encrypted votes whose decryptions were not available in $\bbv$ and individual decrypted votes that were not decrypted by any encrypted vote on $\bbc$ --- should be verifiably identified by the backend servers. And, this must be achieved without leaking any additional information.

Recoverability from output-phase  failures requires identifying which of the electronic vote and the VVPR represents the voter's intent. This may be possible in some cases but not always. For example, if the voter's receipt is available on $\bbr$, then the dispute can be resolved if one can verify in zero-knowledge that the receipt encrypted the electronic vote and not the paper one, or vice versa.

In some cases, the disputes may not be resolvable at all. Consider case $\mathsf{FO}_3$ in Figure \ref{fig:failures} and suppose the receipt is not available. $\mathsf{FO}_3$ may be due to $a)$ the polling officer uploading an encrypted vote not matching the voter's receipt; $b)$ the voter dropping a bogus VVPR into the ballot box; $c)$ a malicious agent altering the VVPRs post-polling; or $d)$ the backend servers not decrypting the uploaded encrypted vote correctly. Different cases point to failures in either the electronic vote or the VVPR and it is not possible to identify the true voter intent. Thus, a conservative way to recover from this situation is to identify the polling booth where the dispute may have originated and conduct only a local re-election at this booth. This must be done without revealing polling booth statistics of at least the uncorrupted polling booths. 

The required action in all the above cases can be reduced to the backend proving in zero-knowledge that an encrypted vote corresponds to one of a set of decrypted votes (a \emph{distributed ZKP of set-membership} \cite{traceable-mixnets}), or that a clear-text vote is a decryption of one of a set of encrypted votes (a \emph{distributed ZKP of reverse set-membership} \cite{traceable-mixnets}).

Cast-as-intended failures may typically happen in two ways. First, ballots may be malformed. Protection against this threat requires a separate audit of a statistically significant sample of ballots before vote casting. Recoverability additionally requires ballot audits to be performed per polling booth. Second, ballots or receipts may be marked incorrectly. In dual voting systems based on hand-marked ballots, the voter may mark the encrypted and the VVPR parts differently, leading to failures. Although this is easily detected and invalidated during VVPR audit, fixing accountability may be difficult and hence voters may do this deliberately to discredit the election. In systems based on ballot marking devices (BMD), such voter errors are avoided but a dispute may be raised that the ballot marking is not according to the voter's choice. Such a dispute between a man and a machine is unresolvable and the only recourse is to allow the voter to revote. This may however cause a deadlock, which can only be resolved through a social process. Still, a BMD should be a preferred option for dual voting since it minimises voter-initiated errors.

\section{Formalisation}
\label{formalisation}

We now formalise the requirements outlined in the previous section. Given a positive integer $x$, let $[x]$ denote the set $\{1,\dots,x\}$. We consider a dual-voting protocol involving $\alpha$ candidates, $n$ voters $\voters$, $\tau$ ballot generators $\gens$, $\ell$ polling booths consisting of polling officers $\pos$, BMDs $\devs$ and physical ballot boxes $\bboxes$, $m$ backend servers $\mixers$, and an auditor $\auditor$. We also assume existence of a public bulletin board where lists $\bbr$, $\bbc$, $\bbv$ and $\bbp$ are published. We consider a protocol structure $(\setup$, $\ballotgen$, $\cast$, $\tally$, $\ballotaudit$, $\receiptaudit$, $\audit)$ where:
\begin{itemize}[leftmargin=*]
	\item $\setup$ is a protocol involving $\gens$, $\pos$ and $\mixers$ to generate public/private key pairs and other public election parameters.
	\item $\ballotgen$ is a protocol involving $\gens$ to securely print a sealed ballot given a booth identifier $j \in [\ell]$.
	\item $\cast$ is the vote casting protocol involving $\voter{i}$ and $\po{j}$, $\dev{j}$, $\bbox{j}$ at booth $j \in [\ell]$ assigned to $\voter{i}$. $\voter{i}$'s input is its intended vote $v$ and a ballot $b$. The protocol outputs a voter receipt $r$, an encrypted vote $c$ and a VVPR $p$ such that $p$ gets dropped in ballot box $\bbox{j}$, $\voter{i}$ takes $r$ home and $\po{j}$ uploads $c$ on $\bbc$. The voter may or may not publish $r$ on $\bbr$. The VVPR is published on $\bbp$ after aggregating VVPRs from all the booths.
	\item $\tally$ is the vote processing/tallying protocol involving $\mixers$ where they take as input the encrypted votes $(\vc{c}{i})_{i \in [n]}$ published on $\bbc$, permute and decrypt them and publish a list $(\vc{v}{i}')_{i \in [n]}$ of decrypted votes on $\bbv$.
	\item $\ballotaudit$ is a protocol involving $\auditor$ and $\po{j}$ executed at each booth $j$ to verify if ballots at booth $j$ are well-formed.
	\item $\receiptaudit$ is a protocol involving $\auditor$ and the voters to verify that voter receipts at booth $j$  match those uploaded on list $C$.
	\item $\audit$ is a protocol involving $\auditor$, $\mixers$ and $\gens$ to verify whether the electronic and paper tallies are correct and narrow down errors if not. It takes as input all published lists $(\bbr,\bbc,\bbv,\bbp)$ and lets $\auditor$ output a tuple $(J^*, V^*)$ where $J^*$ denotes the set of booths that contributed potentially outcome-changing failures and $V^*$ denotes the set of votes from booths not in $J^*$ ($\auditor$ may also be aborted). The expected usage of the $(J^*,V^*)$ output is that in case of failures/disputes, the election could be rerun at the booths in $J^*$ and the rerun results could be merged with the recovered partial tally from $V^*$ to obtain the complete election tally. Results are announced to the general public only after $\audit$ has finished.
\end{itemize}
Note that although the above audits are performed by different auditors (even voters) at different times and places, we simplify  by representing all the auditors by $\auditor$. 

Let $\epsilon_{\mathsf{b}}$ denote the probability that $\mathsf{BallotAudit}$ passes at some booth $j$ yet a receipt from the booth does not encrypt the voter's intent correctly, and $\epsilon_{\mathsf{r}}$ denote the probability that $\mathsf{ReceiptAudit}$ passes for booth $j$ yet a receipt from the booth is not uploaded correctly. Further, let $R^*\subseteq R$, $C^*\subseteq C$ and $P^*\subseteq P$ respectively denote receipts, encrypted votes and VVPRs from booths not in $J^*$.  Finally, let \emph{failures} in a tuple $(R,C,V,P)$ be as defined in Figure \ref{fig:failures} with the added condition that if a receipt or encrypted vote from a booth fails with input-phase or cast-as-intended failures, then all receipts and encrypted votes from that booth are considered as failures.

Definition \ref{def:recoverability} models our notion of recoverability parametrised by probabilities $\epsilon_{\mathsf{b}}$ and $\epsilon_{\mathsf{r}}$ denoting the effectiveness of ballot and receipt audits. The case when $J^*$ is empty denotes that no rerun is required at any booth, either because the election ran completely correctly, or because the number of failures are small compared to the reported winning margin. When non-empty, $J^*$ should exactly be the set of booths where re-run is required because of failures that may affect the final outcome and votes $V^*$ must be consistent with receipts, encrypted votes and VVPRs from booths not in $J^*$. 

Note that the auditor is allowed to abort the $\audit$ protocol, since if the mix-servers and the ballot generators holding the election secrets do not cooperate, then recovery cannot happen. This is not an issue because unlike polling booth failures, these failures are centralised and non-cooperation directly puts the blame on these entities.

\begin{definition}[Recoverability] 
	\label{def:recoverability}
    A voting protocol $(\setup$, $\ballotgen$, $\cast$, $\tally$, $\ballotaudit$, $\receiptaudit$, $\audit)$ is \emph{recoverable} by the audit protocols if 
    for all polynomially bounded adversaries corrupting $\gens$, $\mixers$, $\pos$, $\devs$ and $\bboxes$ such that $\auditor$ outputs a tuple $(J^*, V^*)$ and does not abort, the following conditions hold true with probability only negligibly smaller than $1 - \ell(\epsilon_{\mathsf{b}} + \epsilon_{\mathsf{r}})$:
	\begin{itemize}[leftmargin=*]
		\item if $J^*$ is empty, then the number of failures in $(R^*, C^*, V^*, P^*)$ is less than the reported winning margin computed from $\bbv$; and
		\item if $J^*$ is non-empty, then $(R^*, C^*, V^*, P^*)$ does not contain any failures and $J^*$ is exactly the set of booths that contributed some failing receipt in $\bbr$, some failing encrypted vote in $\bbc$, or some failing VVPR in $\bbp$.
	\end{itemize}
\end{definition}

Definition \ref{def:secrecy-mod-recoverability} models that in the presence of the $\audit$ protocol, the standard vote secrecy guarantee is maintained except that polling booth statistics of the booths contributing some failing items are revealed. This is generally an unavoidable tradeoff. 

\begin{definition}[Vote Secrecy with Recoverability] 
	\label{def:secrecy-mod-recoverability}
	A voting protocol $(\setup$, $\ballotgen$, $\cast$, $\tally$, $\ballotaudit$, $\receiptaudit$, $\audit)$ \emph{protects vote secrecy with recoverability} if no polynomially bounded adversary controlling the auditor $\auditor$, $\pos$, $\devs$, $(\gen{t})_{t \in [\tau] \setminus \{t^*\}}$ for some $t^* \in [\tau]$, $(\mixer{k})_{k \in [m] \setminus \{k^*\}}$ for some $k^* \in [m]$, and $(\voter{i})_{i \in [n] \setminus \{i_0,i_1\}}$ for some $i_0,i_1 \in [n]$ can distinguish between the following two worlds except with negligible probability:
	\begin{itemize}[leftmargin=*]
		\item \textbf{(World 0)} $\voter{i_0}$ votes $v_0$ at booth $j_0$ and $\voter{i_1}$ votes $v_1$ at booth $j_1$, and 
		\item \textbf{(World 1)} $\voter{i_0}$ votes $v_1$ at booth $j_0$ and $\voter{i_1}$ votes $v_0$ at booth $j_1$, 
	\end{itemize}
    where $v_0,v_1$ are any two valid votes and for each failure from booth $j_0$, the adversary must create an identical failure (same failure type and affected vote) from booth $j_1$.
\end{definition} 

\section{The \emph{OpenVoting} Protocol}
\label{sec:protocol}

\subsection{Preliminaries}

\noindent \textbf{Notation.} Let $\Gone, \Gtwo, \GT$ denote cyclic groups of prime order $q$ ($q\gg \alpha,m,n,\ell$) such that they admit an efficiently computable bilinear map $e: \Gone \times \Gtwo \rightarrow \GT$. We assume that the $n$-Strong Diffie Hellman (SDH) assumption \citep{boneh-boyen} holds in $(\Gone, \Gtwo)$, the decisional Diffie-Hellman (DDH) and the discrete logarithm (DL) assumptions hold in $\Gone$, and that generators $g_1, h_1 \in \Gone$ are chosen randomly (say as the output of a hash function) so that nobody knows their mutual discrete logarithm. \\

\noindent \textbf{Traceable Mixnets \citep{traceable-mixnets}.} Traceable mixnets extend traditional mixnets \citep{mixnet-sok} to enable the distributed ZKPs of set membership mentioned in Section \ref{sec:requirements}. Thus, we use them as our cryptographic backend. In traceable mixnets, the backend servers, often also called \emph{mix-servers}, can collectively prove answers to the following queries in zero-knowledge:
\begin{itemize}[leftmargin=*]
	\item TraceIn: whether a ciphertext $c$ (from the mixnet's input ciphertext list) encrypts a value in a subset of output plaintexts (denoted as $(\vc{v}{i}')_{i \in I'}$ for some $I' \subseteq [n]$). 
	\item TraceOut: whether a plaintext $v$ (from the mixnet's output plaintext list) is encrypted in one of a subset of input ciphertexts (denoted as $(\vc{c}{i})_{i \in I}$ for some $I \subseteq [n]$.). 
\end{itemize}
There are also batched versions of these queries called BTraceIn and BTraceOut, which prove multiple TraceIn and TraceOut queries together. 

Formally, a traceable mixnet $\tm$ is a protocol between a set of senders $S_1,\dots,S_n$, a set of mix-servers $(\mixer{k})_{k \in [m]}$ and a querier $\querier$ and consists of algorithms/sub-protocols $(\keygen, \enc, \mix, \btrin, \btrout)$ where:
\begin{itemize}[leftmargin=*]
	\item $\keygen$ is a distributed key generation protocol involving $(\mixer{k})_{k \in [m]}$ that outputs a mixnet public key $\mpk$ and secret keys $\msk{k}$ for each mix-server $\mixer{k}$.
	\item $\enc$ is the encryption algorithm that a sender $S_i$ uses to create a ciphertext $\vc{c}{i}$ encrypting its secret input $\vc{v}{i}$ against $\mpk$.
	\item $\mix$ is the mixing protocol involving $(\mixer{k})_{k \in [m]}$ that takes as input the list of ciphertexts $(\vc{c}{i})_{i \in [n]}$ uploaded by $(S_i)_{i \in [n]}$ and outputs a list of permuted plaintexts $(\vc{v}{i}')_{i \in [n]}$ and a secret witness $\wit{k}$ for each $\mixer{k}$.
	\item $\btrin$ is a protocol involving $(\mixer{k})_{k \in [m]}$ and $\querier$ that takes as input $(\vc{c}{i})_{i \in [n]}$ and $(\vc{v}{i}')_{i \in [n]}$ and index sets $I,I' \subseteq [n]$ (each $\mixer{k}$ additionally uses $\wit{k}$). At the end of the protocol, $\querier$ either outputs the subset of ciphertexts $\{\vc{c}{i}\}_{i \in I}$ that encrypt some plaintext in $\{\vc{v}{i}'\}_{i \in I'}$ or aborts. 
	\item $\btrout$ is a protocol involving $(\mixer{k})_{k \in [m]}$ and $\querier$ that takes exactly the same inputs as $\btrin$. In this case, $\querier$ either outputs the subset of plaintexts $\{\vc{v}{i}'\}_{i \in I'}$ that are encrypted by some ciphertext in $\{\vc{c}{i}\}_{i \in I}$ or aborts. 
\end{itemize}
The \emph{soundness property of traceable mixnets} states that an adversary controlling all $(\mixer{k})_{k \in [m]}$ cannot make $\querier$ output an incorrect set. Their \emph{secrecy property} states that an adversary controlling $(\mixer{k})_{k \in [m] \setminus \{k^*\}}$ for some $k^* \in [m]$, $\querier$ and $(S_i)_{i \in [n] \setminus \{i_0,i_1\}}$ for some $i_0,i_1 \in [n]$ cannot distinguish between a world where $(S_{i_0},S_{i_1})$ respectively encrypt $(v_0,v_1)$ and the world where they encrypt $(v_1,v_0)$, if the $\btrin$ and $\btrout$ query outputs do not leak this information, i.e., if in all $\btrin$ queries, $v_0 \in \{\vc{v}{i}'\}_{i \in I'}$ iff $v_1 \in \{\vc{v}{i}'\}_{i \in I'}$ and in all $\btrout$ queries, $i_0 \in I$ iff $i_1 \in I$. \\

\noindent \textbf{An instantiation of traceable mixnets.} \citep{traceable-mixnets} also provides a concrete instantiation of a traceable mixnet, which we use. In this instantiation, $\mpk$ is of the form $((\pkey{\mixer{k}})_{k \in [m]}$, $\thegpk$, $\thezpk)$, where $\pkey{\mixer{k}}$ is the public key of any IND-CPA secure encryption scheme $\encscheme$, and $\thegpk$ and $\thezpk$ are respectively public keys of $\theg$, the threshold ElGamal encryption scheme \citep{threshold-cryptography} with message space $\Gone$, and $\thez$, the threshold Paillier encryption scheme proposed in \citep{dj-generalisation} with message space $\mathbb{Z}_N$ for an RSA modulus $N$. The secret key $\msk{k}$ for each $\mixer{k}$ consists of the secret key $\skey{\mixer{k}}$ corresponding to $\pkey{\mixer{k}}$ and the $k^{\text{th}}$ shares of the secret keys corresponding to $\thegpk$ and $\thezpk$. Further, $\enc$ on input a value $v \in \Zq$ outputs a ciphertext of the form $(\epsilon, \gamma, (\share{\mathsf{ev}}{k},\share{\mathsf{er}}{k})_{k \in [m]}, \pokcomm, \epsilon_r)$, where
\begin{itemize}
	\item $\epsilon \leftarrow \theze(\thezpk, v)$ is an encryption of $v$ (interpreted as $v \in \mathbb{Z}_N$) under $\thez$,
	\item $\gamma = g_1^vh_1^r$ is a Pedersen commitment \citep{Pedersen} to $v$ in $\Gone$ under randomness $r \in \Zq$,
	\item $\share{\mathsf{ev}}{k} \leftarrow \pkee(\pkey{\mixer{k}}, \share{v}{k})$ is an encryption of a secret share $\share{v}{k}$ of $v$,
	\item $\share{\mathsf{er}}{k} \leftarrow \pkee(\pkey{\mixer{k}}, \share{r}{k})$ is an encryption of a secret share $\share{r}{k}$ of $r$,
	\item $\pokcomm \leftarrow \nizkpk{v,r}{\gamma = g_1^v h_1^r}$ is a noninteractive ZKP of knowledge of the opening of $\gamma$, and 
	\item $\epsilon_r \leftarrow \theze(\thezpk, r)$ is an encryption of $r$ (interpreted as $r \in \mathbb{Z}_N$) under $\thez$.
\end{itemize}
In our protocol, the encrypted votes are encryptions under $\enc$, where we instantiate scheme $\pke$ with the (non-threshold) Paillier encryption scheme \citep{paillier}. We need it for its following homomorphic property: given two Paillier ciphertexts $c_1,c_2$ encrypting messages $m_1,m_2 \in \Zq$ respectively ($m_1,m_2$ interpreted as messages in $\mathbb{Z}_N$), the ciphertext $c_1c_2$ encrypts the message $m_1+m_2 \mod N = m_1+m_2$ if $N>2q$. We also require a public-key digital signature scheme $\pks := (\keygen$, $\sign$, $\verify)$ with the usual existential unforgeability property under chosen message attacks (EUF-CMA).

\subsection{The Proposed Protocol}

\begin{figure*}[t]
	\centering
	\includegraphics[width=0.95\textwidth]{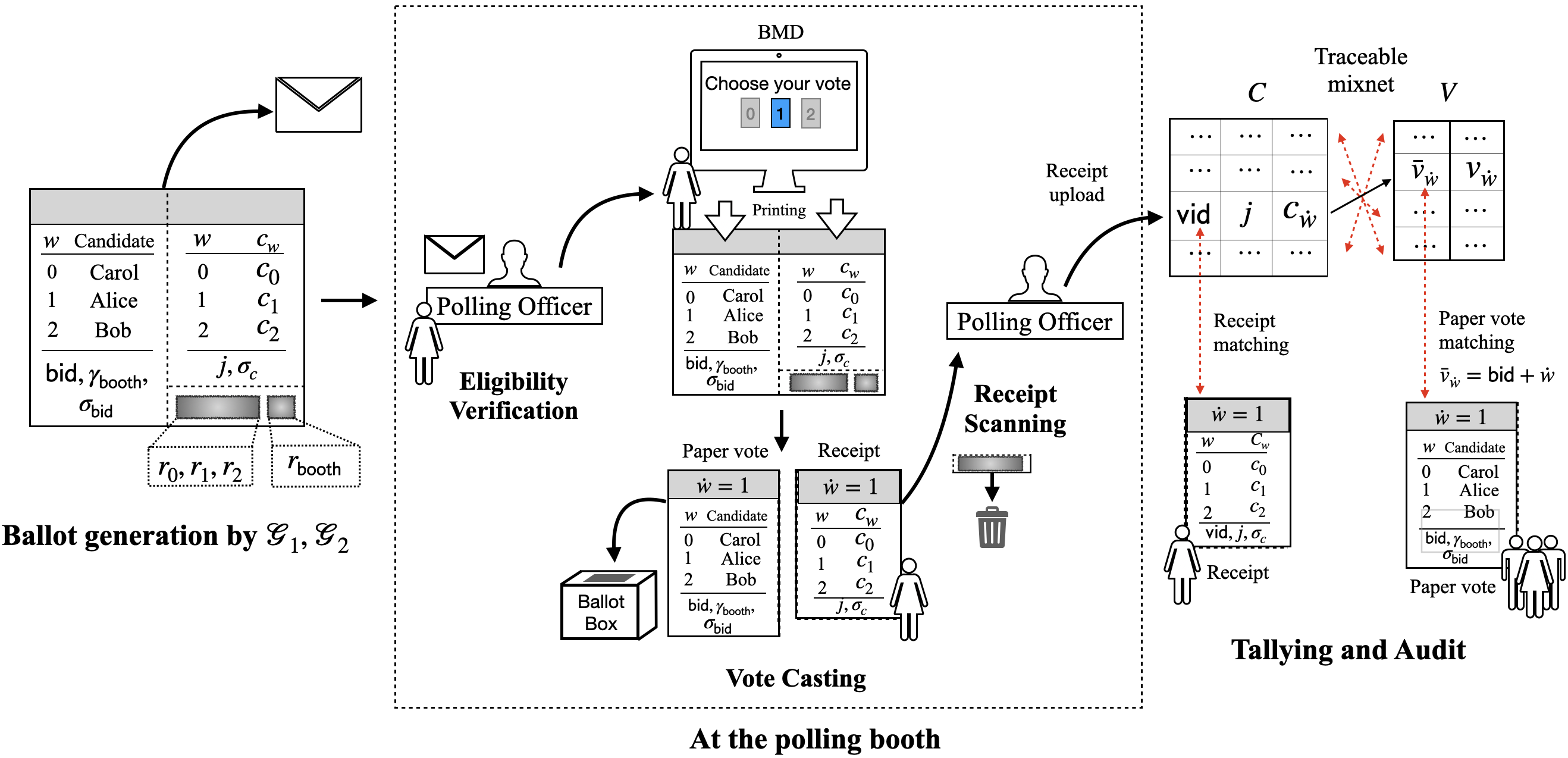}
	\caption{Overview of the \emph{OpenVoting} protocol: $w$ represents the candidate index in the ballot and $\dot{w}$ represents the voter's choice.}
	\label{fig:overview}
\end{figure*}

Figure \ref{fig:overview} depicts the high-level \emph{OpenVoting} protocol. Two ballot generators ($\gen{1}$ and $\gen{2}$) jointly generate sealed ballots to protect voter-vote association from both. Voters use the sealed ballots and a BMD to cast their votes. Each ballot contains two halves. The BMD prints the voter's choice on both halves without learning the vote. The left half becomes the VVPR and is deposited by the voter in a physical ballot box, while the right half becomes the voter receipt. Polling officers scan the voter receipts and upload the encrypted votes to $\bbc$. The encrypted votes are processed by a traceable mixnet backend to produce decrypted votes $V$. Voters can verify their receipts against the encrypted votes, and VVPRs can be matched with the decrypted votes. The tally audit process uses the traceable mixnet's querying mechanism to identify polling booths contributing to failures without leaking additional information. Results are announced only after this audit step. Now we describe the sub-protocols of \emph{OpenVoting} in detail.

\subsubsection{Setup} 
\label{ss:setup}

During the $\setup$ protocol, $\gen{1}$ and $\gen{2}$ generate public/private keys $\pkey{\gen{1}}, \skey{\gen{1}}$ and $\pkey{\gen{2}}, \skey{\gen{2}}$ under $\pks$. Polling officers $(\po{j})_{j \in [\ell]}$ also generate public/private keys $(\pkey{\po{j}}, \skey{\po{j}})_{j \in [\ell]}$ under $\pks$. Mix-servers $(\mixer{k})_{k \in [m]}$ jointly run the $\tm.\keygen$ protocol of the traceable mixnet to generate the mixnet public key $\mpk$ and individual secret keys $(\msk{k})_{k \in [m]}$ for each $(\mixer{k})_{k \in [m]}$. An official candidate list $(\mathsf{cand}_0,\dots,\mathsf{cand}_{\alpha-1})$ is created such that $\mathsf{cand}_a$ denotes the $a^{\text{th}}$ candidate.

\subsubsection{Ballot Design}
\label{ss:ballot-design}

Our ballot (Figure \ref{fig:overview} - left) customises the Scratch \& Vote ballot \citep{scratchandvote} for dual voting and BMD support. It consists of two halves connected by a perforated line. The left half serves as the VVPR, while the right half serves as the voter receipt. These halves are unlinkable after the vote is cast.

The left half includes a randomly drawn ballot identifier $\bid$ from $\mathbb{Z}_q$. It displays a circular rotation of the official candidate list. For each $w \in \{0,\dots,\alpha-1\}$, row $w$ corresponds to the candidate $\mathsf{cand}_{\bid + w \bmod{\alpha}}$. For example, if the official candidate list is $(\text{``Alice''}, \text{``Bob''}, \text{``Carol''})$, and $\bid=302$, the candidate printed on row $w=1$ would be $\mathsf{cand}_{302+1 \bmod 3} = \mathsf{cand}_0$ (i.e., ``Alice'').
The right half contains corresponding encryptions $c_{w}$ obtained by running the $\tm.\enc$ algorithm on input $\bar{v}_w = \bid+w$, except that they do not include the $\rho_{\gamma_w}$ component; this is added during the $\tally$ protocol. We call values $\bar{v}_w \in \mathbb{Z}_{q}$ the \emph{extended votes} and values $v_w := \bar{v}_w \bmod{\alpha} \in [\alpha]$ the \emph{raw votes}.  The randomnesses $r_w$ used in creating encryption $c_w$ are kept secret and placed under a detachable scratch surface on the ballot.

Both halves feature a designated gray area at the top. During the $\cast$ protocol, the BMD prints the voter-selected $w$ in this gray area on both halves. Additionally, the right half includes a polling booth identifier $j$ for the designated polling booth of the ballot, while the left half contains its commitment $\gamma_{\booth} = g^{j}h^{r_{\booth}}$. Randomness $r_{\booth} \xleftarrow{\$} \Zq$ is also put under a separate scratch surface. The commitment $\gamma_{\booth}$ is revealed when the polling booth of a disputed VVPR needs to be identified in the $\audit$ protocol.

Due to the size of encryptions $c_w$ (around 20 KB each \citep{traceable-mixnets}), they may not fit within standard QR codes on the paper ballot. However, conceptually, the actual encryptions could be stored in a backend server, with only a binding hash printed on the ballot. For simplicity, we ignore this complication.

\subsubsection{Ballot Generation}
\label{ss:ballotgen}

During the $\ballotgen$ protocol, a ballot is jointly generated by $\gen{1}$ and $\gen{2}$ to hide the voter-vote association from any one of them (see Figure \ref{fig:ballotgen-protocol}). $\gen{1}$, who selects the ballot secrets, does not learn the encryptions printed on the receipt half and cannot match voters to their ballot secrets, while $\gen{2}$, who creates the receipt half, does not know the ballot secrets.

$\gen{2}$ knows the destination booth $j$ but keeps it hidden from $\gen{1}$ to hide booth-level voting statistics. It generates a commitment $\gamma_{\booth}$ for $j$ and shares it with $\gen{1}$ (lines 1-2), who prints it on the left half of the ballot. $\gen{1}$ generates a secret ballot identifier $\bid$ and signs it (lines 2-3), computes $\bar{v}_w = \bid+w$ for each $w$ and accordingly prints candidate names on the left half and randomnesses $r_w$ under a scratch surface on the right half (lines 6,10). It then sends the partially printed ballot to $\gen{2}$, keeping the left half hidden. This can be done, e.g., by folding the ballot along the perforation line, sealing it and letting $\gen{2}$ print its contents on the \emph{back side} of the right half. It also sends encryptions of each $\bar{v}_w$ under $\tm.\enc$, except the $\rho_{\gamma_w}$ components, to $\gen{2}$ (lines 7-9,12).

$\gen{2}$ re-randomises the obtained commitments/encryptions and homomorphically computes fresh shares of $\bar{v}_w$ and the commitment randomnesses using the additive homomorphism of $\pke$ in $\Zq$ (lines 13-19). It then prints these re-randomised encryptions on the right half of the received ballot and signs them. The re-randomisation ensures that $\gen{1}$ cannot identify the ballot corresponding to a voter from their receipt. The commitment randomness $r_{\booth}$ of $\gamma_{\booth}$ is printed on another scratch surface on the right half.

\begin{figure}
	\centering
	\scalebox{0.8}{
		\begin{tabular}{rll}
			\hline
			1 & $\gen{2}$: & $r_{\booth} \xleftarrow{\$} \Zq$; $\gamma_{\booth} \leftarrow g_1^{j}h_1^{r_{\booth}}$ \\
			2 & $\gen{2}$: & send $\gamma_{\booth}$ to $\gen{1}$ \\ 
			3 & $\gen{1}$: & $\bid \xleftarrow{\$} \Zq$ \\
			4 &            & $\sigma_{\bid} \leftarrow \pks.\sign(\skey{\gen{1}}, \bid)$ \\
			5 & 		   & \textbf{for $w \in \{0,\dots,\alpha-1\}$:} \\
			6 &		       & \quad $\bar{v}_{w} \leftarrow \bid + w$; $r_{w} \xleftarrow{\$} \Zq$;  $\gamma_{w} \leftarrow g_1^{\bar{v}_{w}}h_1^{r_{w}}$ \\
			7 &		       & \quad $\epsilon_{\bar{v}_w} \leftarrow \theze(\thezpk, \bar{v}_w)$; $\epsilon_{r_w} \leftarrow \theze(\thezpk, r_w)$ \comment{interpret $v_w,r_w$ as elements of $\mathbb{Z}_N$} \\
			8 &		       & \quad $(\share{\bar{v}}{k}_{w})_{k \in [m]} \leftarrow \sharealg_{(m,m)}(\bar{v}_{w})$; $(\share{r}{k}_{w})_{k \in [m]} \leftarrow \sharealg_{(m,m)}(r_{w})$ \\
			9 &		       & \quad $(\share{\mathsf{ev}}{k}_{w})_{k \in [m]} \leftarrow (\pke.\enc(\pkey{\mixer{k}}, \share{\bar{v}}{k}_{w_i}))_{k \in [m]}$;  $(\share{\mathsf{er}}{k}_{w_i})_{k \in [m]} \leftarrow (\pke.\enc(\pkey{\mixer{k}}, \share{r}{k}_{w_i}))_{k \in [m]}$ \\
			10 &		   & print ballot's left half ($\mathsf{cand}_{\bar{v}_{w} \bmod \alpha}$ on row $w$) and $(r_{w})_{w \in \{0,\dots,\alpha-1\}}$ under a scratch surface as per Fig. \ref{fig:overview} - left \\
			11 &		   & send the ballot to $\gen{2}$ with its left half sealed \\
			12 &		   & send $(\epsilon_{\bar{v}_w}, \gamma_{w}, (\share{\mathsf{ev}}{k}_{w}, \share{\mathsf{er}}{k}_{w})_{k \in [m]}, \epsilon_{r_w})_{w \in \{0,\dots,\alpha-1\}}$ to $\gen{2}$ electronically \\
			13 & $\gen{2}$: & \textbf{for $w \in \{0,\dots,\alpha-1\}$:} \\
			14 & 		   & \quad $r_{w}' \xleftarrow{\$} \Zq$; $\gamma_{w}' \leftarrow \gamma_{w} h_1^{r_{w}'}$ \\
			15 &		   & \quad $\epsilon'_{\bar{v}_w} \leftarrow \thezre(\thezpk, \epsilon_{\bar{v}_w})$; $\epsilon'_{r_w} \leftarrow \thezre(\thezpk, \epsilon_{r_w})$ \quad \comment{$\thezre(\thezpk, \epsilon) = \epsilon\theze(\thezpk, 0)$}\\
			16 &		   & \quad $(\share{v'}{k}_{w})_{k \in [m]} \leftarrow \sharealg_{(m,m)}(0)$; $(\share{r'}{k}_{w})_{k \in [m]} \leftarrow \sharealg_{(m,m)}(r_{w}')$ \\
			17 &		   & \quad $(\share{\mathsf{ev}'}{k}_{w})_{k \in [m]} \leftarrow (\share{\mathsf{ev}}{k}_{w} \cdot \pke.\enc(\pkey{\mixer{k}}, \share{v'}{k}_{w}))_{k \in [m]}$ \\
			18 &		   & \quad $(\share{\mathsf{er}'}{k}_{w})_{k \in [m]} \leftarrow (\share{\mathsf{er}}{k}_{w} \cdot \pke.\enc(\pkey{\mixer{k}}, \share{r'}{k}_{w}))_{k \in [m]}$ \\				   
			19 &		   & \quad $c_w := (\epsilon'_{\bar{v}_w}, \gamma'_w, (\share{\mathsf{ev}'}{k}_{w}, \share{\mathsf{er}'}{k}_{w})_{k \in [m]}, \epsilon'_{r_w})$ \\
			20 &		   & $\sigma_c \leftarrow \pks.\sign(\skey{\gen{2}}, H((c_w)_{w \in \{0,\dots,\alpha-1\}}))$, where $H$ is a hash function \\
			21 &		   & print ballot's right half and $r_{\booth}$ under another scratch surface as per Fig. \ref{fig:overview} - left \\
			22 &		   & store $(j,r_{\booth})$ indexed by $\gamma_{\booth}$ \\
			\hline
		\end{tabular}
	}
	\caption{The $\ballotgen$ protocol for generating a ballot for booth $j$ known only to $\gen{2}$.}
	\label{fig:ballotgen-protocol}
\end{figure}

\subsubsection{Vote Casting}
\label{ss:vote-casting}

The $\cast$ protocol for voter $\voter{i}$ at booth $j$ is as follows (Figure \ref{fig:overview} - center):
\begin{itemize}[leftmargin=*]
	\item \emph{Ballot pick-up and eligibility verification:} $\voter{i}$ picks up a random sealed ballot from a set of ballots kept at the polling booth. The polling officer $\po{j}$ verifies $\voter{i}$'s eligibility in the presence of polling agents and allows $\voter{i}$ to proceed to a private room containing a BMD $\dev{j}$. \label{enum:vote-casting-1}
	\item \emph{Vote casting:} $\voter{i}$ feeds the top gray region of the ballot to $\dev{j}$ and presses a button on the onscreen display to select $w$ corresponding to her preferred candidate. We denote the voter's chosen $w$ as $\dot{w}$. $\dev{j}$ can only access the top gray region for printing and cannot read any part of the ballot (it should not have any attached scanner or camera). $\dev{j}$ prints $\dot{w}$ on both the left and the right halves of this gray region.
	
	$\voter{i}$ needs to verify that indeed her intended choice is printed on both the halves. If satisfied, $\voter{i}$ separates the left half of the marked ballot (the VVPR), folds it and drops it into a physical ballot box $\bbox{j}$ kept near $\po{j}$ such that $\po{j}$ can verify that the voter dropped an official VVPR. The right half (the receipt) is given to $\po{j}$ for scanning. If not satisfied, $\voter{i}$ shreds the marked ballot and raises a dispute. In this case, $\voter{i}$ is allowed to re-vote. Note that the vote casting phase can also completely avoid the BMD and require the voter to hand-mark the two ballot halves, but this design is prone to more voter errors (see Section \ref{sec:requirements}).
	\item \emph{Receipt scanning:} $\po{j}$ checks that the scratch surface on $\voter{i}$'s receipt is intact, i.e., the ballot secrets are not compromised, and shreds the scratch region in front of $\voter{i}$. From the scanned receipt, $\po{j}$ extracts $c_{\dot{w}}$ and uploads $(\vid_i, j, c_{\dot{w}})$ to $\bbc$, along with $\sigma_{\vid_i} \leftarrow \pks.\sign(\skey{\po{j}}, (\vid_i, j, c_{\dot{w}}))$. $\po{j}$ also affixes $\vid_i$ to $\voter{i}$'s receipt, stamps it for authenticity, and returns it to $\voter{i}$. 
\end{itemize}

\noindent \emph{Chain voting and randomisation attacks}. With minor modifications, these sophisticated coercion attacks can also be handled. For chain voting, $\po{j}$ can stamp a serial number on the receipt half of the sealed ballot after identity verification to prevent the use of rogue ballots. This number is matched before accepting the voter's receipt. Under a \emph{randomisation attack}, voters may be asked to choose a fixed $\dot{w}$, thereby randomising their votes. To counter this, voters should be allowed to choose their ballots in a private room. The ballot cover should contain a detachable slip showing the candidate order, allowing coerced voters to choose a ballot so that they can vote for their preferred candidate while producing the $\dot{w}$ satisfying the coercer. Before proceeding to $\po{j}$, the voter should detach the slip.

\subsubsection{Vote Tallying}
\label{sss:vote-processing}

Post polling, $(\mixer{k})_{k \in [m]}$ process the tuples $\{(\vid_i, j_i, c_{i})\}_{i=0}^{n-1}$ uploaded on $\bbc$ by $\pos$, where $c_i$ denotes $c_{\dot{w}}$ for the $i^{\text{th}}$ voter (Figure \ref{fig:overview} - right). $(\mixer{k})_{k \in [m]}$ proceed as per Figure \ref{fig:tally-protocol} where they first add the ${\pokcomm}_i$ components to the encryptions $c_{i}$ by engaging in a distributed NIZK proof of knowledge (lines 2-9) and then processing $(\vc{c}{i})_{i \in [n]}$ through the traceable mixnet's $\mix$ protocol (line 13). At the end, the permuted extended votes $(\vc{\bar{v}}{i}')_{i \in [n]}$ are obtained from which the raw votes are computed (line 14). Both extended and raw votes are published on $\bbv$.

The VVPRs from each polling booth's ballot box are collected and mixed in a central facility. VVPRs are revealed to the public only after this mixing phase and post audit, to avoid leaking polling booth-level voting statistics. A VVPR containing ballot identifier $\bid$ and voter choice $\dot{w}$ can be matched with the corresponding decrypted vote by computing $\bid+\dot{w}$, finding it on $\bbv$ and checking if the corresponding raw vote matches the candidate name printed on the $\dot{w}^{\text{th}}$ row on the VVPR.

\begin{figure}
	\centering
	\scalebox{0.8}{
		\begin{tabular}{rll}
			\hline
			1 & $(\mixer{k})_{k \in [m]}$: 
			                 & \textbf{for $i \in [n]$:} \\
			2 &              & \quad $\vcs{\bar{v}}{i}{k} \leftarrow \pked(\skey{\mixer{k}}, \vcs{\mathsf{ev}}{i}{k})$ \quad \comment{decryption under $\pke$}\\
			3 &              & \quad $\vcs{r}{i}{k} \leftarrow \pked(\skey{\mixer{k}}, \vcs{\mathsf{ev}}{i}{k})$ \\
			4 &              & \quad // Generate a distributed NIZK PoK $\rho_{\gamma_i}$ of the opening of $\gamma_i$ using shares $(\vcs{\bar{v}}{i}{k}, \vcs{r}{i}{k})_{k \in [m]}$ \\
			5 &              & \quad $\vcs{r}{v_i}{k}, \vcs{r}{r_i}{k} \xleftarrow{\$} \Zq$; $\vcs{a}{i}{k} \leftarrow g_1^{\vcs{r}{v_i}{k}}h_1^{\vcs{r}{r_i}{k}}$; publish $\vcs{a}{i}{k}$. \\
	        6 &              & \quad $\mathsf{c}_i \leftarrow H(\gamma_i \Vert \prod_{k \in [m]} \vcs{a}{i}{k})$; $\vcs{z}{\bar{v}_i}{k} \leftarrow \vcs{r}{v_i}{k} - \vcs{\bar{v}}{i}{k}\mathsf{c}_i$; $\vcs{z}{r_i}{k} \leftarrow \vcs{r}{r_i}{k} - \vcs{r}{i}{k}\mathsf{c}_i$; publish $\vcs{z}{\bar{v}_i}{k},\vcs{z}{r_i}{k}$. \\
	        7 &              & \quad $\vc{\pokcomm}{i}:= (a_i,\mathsf{c}_i,(z_{\bar{v}_i},z_{r_i}))$ $\leftarrow$ $(\prod_{k \in [m]} \vcs{a}{i}{k}$, $H(\gamma_i \Vert \prod_{k \in [m]} \vcs{a}{i}{k})$, $(\sum_{k \in [m]}\vcs{z}{\bar{v}_i}{k}, \sum_{k \in [m]}\vcs{z}{r_i}{k}))$. \\
			8 &			     & \quad // $\vc{\pokcomm}{i}$ can be verified by checking if $\mathsf{c}_i \stackrel{?}= H(\gamma_i \Vert a_i)$ and $\gamma_i^{\mathsf{c}_i}g_1^{z_{\bar{v}_i}}h_1^{z_{r_i}} \stackrel{?}= a_i$. \\
			9 &			     & \quad update $\vc{c}{i}$ by inserting $\vc{\pokcomm}{i}$ into it  \\
			10 &			 & \textbf{endfor} \\
			11 &			 & // Mixing protocol to generate permuted \emph{extended} votes \\
			12 &			 & // Each $\mixer{k}$ gets secret input $\msk{k}$ and secret output $\wit{k}$ (see Section \ref{sss:traceable-mixnets}) \\  
			13 & $(\mixer{k})_{k \in [m]}$:			 
			                 & $(\vc{\bar{v}}{i}')_{i \in [n]}, (\so{\mixer{k}}{\wit{k}})_{k \in [m]} \leftarrow \tm.\mix(\mpk, (\vc{c}{i})_{i \in [n]}, (\si{\mixer{k}}{\msk{k}})_{k \in [m]})$ \\
			14 &			 & $(\vc{v}{i}')_{i \in [n]} \leftarrow (\vc{\bar{v}}{i}' \bmod \alpha)_{j \in [n]}$ \\
			15 &			 & publish $(\vc{\bar{v}}{i}')_{i \in [n]}$, $(\vc{v}{i}')_{i \in [n]}$ to $\bbv$; $\mixer{k}$ stores $\wit{k}$ \\
			\hline
		\end{tabular}
	}
	\caption{The $\tally$ protocol involving $(\mixer{k})_{k \in [m]}$ on input $\mpk, (\vc{c}{i})_{i \in [n]}$ and $\mixer{k}$'s input $\msk{k}$ containing $\skey{\mixer{k}}$.}
	\label{fig:tally-protocol}
\end{figure}

\subsubsection{Ballot and Receipt Audits}
\label{ss:ballot-audits}

In the $\ballotaudit$ protocol, a statistically significant number of ballots at each polling booth must be audited to keep the probability $\epsilon_{\mathsf{b}}$ of a cast-as-intended failure (see Section \ref{formalisation}) small. Ballot audits can happen before, during or after polling, and even be initiated by voters. When auditing a ballot, its sealed cover is opened and secrets under its scratch surfaces are revealed. For each $w = 0 \dots \alpha-1$, it is checked that encryption $c_w$ is created correctly on message $\bid+w$ using $r_w$ and the candidate name printed at row $w$ is $\mathsf{cand}_{\bid+w \bmod{\alpha}}$, where $\bid$ is looked up from the left half and $r_w$ from the scratch surface. Further, it is checked that $\gamma_{\booth}\stackrel{?}=g_1^{j}h_1^{r_{\mathsf{booth}}}$, where $j$ is the audited booth's identifier and $r_{\mathsf{booth}}$ is obtained from the scratch surface, and that signatures by $\gen{1}, \gen{2}$ verify. Since the secrets of audited ballots are revealed, audited ballots cannot be used for vote casting and must be spoiled. 

Similarly, in the $\receiptaudit$ protocol, a statistically significant number of voter receipts from each polling booth must be checked for their existence on list $\bbc$ to keep $\epsilon_{\mathsf{r}}$ small. All audited receipts should be uploaded to $\bbr$ to aid audit and recovery. 

\subsubsection{Tally Audit}
\label{sss:audit}

\begin{figure}
	\centering
	\scalebox{0.8}{
		\begin{tabular}{rl}
			\hline
			1 & $J_{\text{FC}} \leftarrow \{j \in [\ell] \mid \ballotaudit \text{ fails at booth } j\}$ \\
			2 & $\R_{\text{FI}} \leftarrow \{r \in \bbr \mid r \text{ fails against } \bbc \text{ under } \mathsf{FI}_1, \mathsf{FI}_2 \}$; $\R_{\text{FI}} \leftarrow \{(\vid,j,c) \in \bbr \mid (\vid',j,c') \in \R_{\text{FI}} \}$  \\
			3 & $\C_{\text{FI}} \leftarrow \{c \in \bbc \mid c \text{ fails against } \bbr \text{ under } \mathsf{FI}_2 \}$; $\C_{\text{FI}} \leftarrow \{(\vid,j,c) \in \bbc \mid (\vid',j,c') \in \C_{\text{FI}} \vee j \in J_{\text{FC}} \}$ \\
			4 & $\C_{\text{FM}} \leftarrow \{\vc{c}{i}\}_{i \in [n_{\mathsf{c}}]} \setminus \btrin(\mpk, (\vc{c}{i})_{i \in [n_{\mathsf{c}}]}, (\vc{\bar{v}}{i}')_{i \in [n_{\mathsf{v}}]}, [n_{\mathsf{c}}], [n_{\mathsf{v}}], (\si{\mixer{k}}{\msk{k}, \wit{k}})_{k \in [m]}, \si{\auditor}{})$ \\
			5 & $\V_{\text{FM}} \leftarrow \{\vc{v}{i}'\}_{i \in [n_{\mathsf{v}}]} \setminus \btrout(\mpk, (\vc{c}{i})_{i \in [n_{\mathsf{c}}]}, (\vc{\bar{v}}{i}')_{i \in [n_{\mathsf{v}}]}, [n_{\mathsf{c}}], [n_{\mathsf{v}}], (\si{\mixer{k}}{\msk{k}, \wit{k}})_{k \in [m]}, \si{\auditor}{})$ \\
			6 & $\V_{\text{FO}} \leftarrow \{\bar{v} \in \bbv \mid \bar{v} \text{ fails against } \bbp \text{ under } \mathsf{FO}_1, \mathsf{FO}_3, \mathsf{FO}_4 \} $ \\
			7 & $\Pa_{\text{FO}} \leftarrow \{p \in \bbp \mid p \text{ fails against } \bbv \text{ under } \mathsf{FO}_2, \mathsf{FO}_3, \mathsf{FO}_5 \} $ \\
			8 & $\R_{\text{F}} \leftarrow \R_{\text{FI}}$; $\C_{\text{F}} \leftarrow \C_{\text{FI}} \cup \C_{\text{FM}}$; $\V_{\text{F}} \leftarrow \V_{\text{FM}} \cup \V_{\text{FO}}$; $\Pa_{\text{F}} \leftarrow \Pa_{\text{FO}}$ \\
			9 & \textbf{if} $|\R_{\text{F}}| + |\C_{\text{F}}| + |\V_{\text{F}}| + |\Pa_{\text{F}}| < $ \text{winning margin calculated from $\bbv$:} \\
			10 & \quad $J^* \leftarrow \emptyset$; $\V^* \leftarrow \bbv$  \\
			11 & \textbf{else:} \\
			12 & \quad $\badbooths_{r} \leftarrow \{ j \mid (\vid,j,c) \in \R_{\text{F}} \}$; $\badbooths_{c} \leftarrow \{ j \mid (\vid,j,c) \in \C_{\text{F}} \}$ \\
			13 & \quad $\badbooths_{p} \leftarrow \{ j \mid \gen{2} \text{ supplies } (j,r_{\booth}) \text{ to } $\auditor$ \text{ for } \gamma_{\booth} \text{ printed on some } p \in \Pa_{\text{F}} \text{ s.t. } \gamma_{\booth} = g_1^{j}h_1^{r_{\booth}} \}$ \\
			14 & \quad $\badbooths_{v} \leftarrow \emptyset$ \\
			15 & \quad \textbf{for $j$ in $[\ell]$:} \\
			16 & \quad \quad \comment{\iindices{j} denotes indices of booth $j$'s entries in $\bbc$; $\jindices{\V_{\text{F}}}$ denotes indices of $\V_{\text{F}}$ entries on $\bbv$ } \\
			17 & \quad \quad $\V_{\text{F}_j} \leftarrow \btrout(\mpk, (\vc{c}{i})_{i \in [n_{\mathsf{c}}]}, (\vc{v}{i}')_{i \in [n_{\mathsf{v}}]}, \iindices{j}, \jindices{\V_{\text{F}}}, (\si{\mixer{k}}{\msk{k}, \wit{k}})_{k \in [m]}, \si{\auditor}{})$ \\
			18 & \quad \quad \textbf{if}~ $\V_{\text{F}_j} \neq \emptyset$\textbf{:} $\badbooths_{v} \leftarrow \badbooths_{v} \cup \{j\}$ \\
			19 & \quad $J^* \leftarrow \badbooths_{r} \cup \badbooths_{c} \cup \badbooths_{p} \cup \badbooths_{v}$ \\ 
			20 & \quad \comment{$\iindices{\goodbooths}$ denotes indices of booths outside $J^*$ in $\bbc$; $\jindices{\V \setminus \V_{\text{F}}}$ denotes indices of entries outside $\V_{\text{F}}$ on $\bbv$ } \\
			21 & \quad $\V^* \leftarrow \btrout(\mpk, (\vc{c}{i})_{i \in [n_{\mathsf{c}}]}, (\vc{v}{i}')_{i \in [n_{\mathsf{v}}]}, \iindices{\goodbooths}, \jindices{\V \setminus \V_{\text{F}}}, (\si{\mixer{k}}{\msk{k}, \wit{k}})_{k \in [m]}, \si{\auditor}{})$ \\
			22 & \textbf{return $J^*, \V^*$} \\
			\hline
		\end{tabular}
	}
	\caption{The $\audit$ protocol involving $\auditor$, $(\mixer{k})_{k \in [m]}$ and $\gen{2}$ with public input $(R,C,V,P)$, each $\mixer{k}$'s input its mixnet secret key $\msk{k}$ and witness $\wit{k}$ output by the traceable mixnet during the $\tally$ protocol, and $\gen{2}$'s input being $(j,r_{\booth})$ stored indexed by $\gamma_{\booth}$ at the end of the $\ballotgen$ protocol.}
	\label{fig:audit-protocol}
\end{figure}

Our tally audit protocol (see Figure \ref{fig:audit-protocol}) depends on BTraceIn and BTraceOut queries of a traceable mixnet (see Section \ref{sss:traceable-mixnets}). Given $(R,C,V,P)$, first, all input-phase failures are marked (lines 1-3). Here, as per the discussion in Section \ref{sec:requirements}, we mark all receipts/encrypted votes from a booth as failed if any one of them fails and the encrypted votes as failed if the ballot audit at that booth failed. For marking mixing phase failures on $\bbc$ and $\bbv$, we run the BTraceIn/BTraceOut queries against the complete set of entries on $\bbv$ and $\bbc$ respectively (lines 4-5). Output-phase failures are marked by comparing the VVPRs with the decrypted extended votes (lines 6-7). 

If the total number of failures is less than the winning margin, then $J^* = \emptyset$ and $V^*=V$ are reported, signalling that no rerun is required (lines 9-10). Otherwise, polling booths contributing all the failing items are identified: for receipts and encrypted votes, the booth identifiers directly exist on $\bbr$ and $\bbc$ (line 12); for VVPRs without an electronic entry, they are identified by asking $\gen{2}$ to open the opening of $\gamma_{\booth}$ printed on the VVPR (line 13); for decrypted votes, a $\btrout$ query against the set of ciphertexts cast at a booth $j$ is run for all booths $j \in [\ell]$ (lines 14-18). The set of all such booths is reported in $J^*$ (line 19). The decrypted votes $V^*$ contributed by the good booths are obtained by running another $\btrout$ query against the entries on $\bbc$ contributed by booths outside of $J^*$ (line 21).

\subsubsection{Recovery} 

The suggested recovery is to rerun the election only on booths in $J^*$ and later merge this tally with the tally reported in $V^*$. However, if $J^*$ is small, one can also consider rerunning on a few randomly selected good booths too, to avoid specialised targeting of the booths in $J^*$. Further, the general approach of TraceIn/TraceOut queries can also support other recovery options for dual voting systems. For example, one can immmediately recover from case $\mathsf{FO}_3$ if a TraceOut query is run for the decrypted vote against the set of encrypted votes that successfully matched with voter receipts. If the answer is yes, then it provides solid evidence that the VVPR is wrong, without leaking any additional information. A similar query run for the VVPR provides solid evidence that the electronic vote was wrong. Of course, what queries to allow must be carefully decided depending on the situation to best optimise the recoverability-secrecy tradeoff.

\section{Security Analysis}

\begin{theorem}
	Under the DL assumption in $\Gone$, the $n$-SDH assumption in $(\Gone, \Gtwo)$ \citep{boneh-boyen} and the EUF-CMA security of $\pks$, the \emph{OpenVoting} protocol is recoverable as per Definition \ref{def:recoverability}.
\end{theorem}
\begin{proof}[Proof (Sketch)]
    We focus on the event that for each booth, $\ballotaudit$ passing implies that all receipts correctly captured voter intents and $\receiptaudit$ passing implies that all receipts were correctly uploaded. This event happens with probability $1-\ell(\epsilon_{\mathsf{b}}+\epsilon_{\mathsf{r}})$.
	
	Let $J^*,V^*$ be $\auditor$'s output in the $\audit$ protocol. From Figure \ref{fig:audit-protocol}, we consider the two cases: first when the branch on line 8 is taken and the second when it is not taken. In the first case, $J^* = \emptyset$ and thus we must show that the number of failures in $(R^*,C^*,V^*,P^*)$ is less than the winning margin, where $R^*=R$, $C^*=C$ and $P^*=P$ for $J^*= \emptyset$ and $V^* = V$ by line 10. By the condition on line 9, the number of \emph{reported} failures is less than the winning margin. By the soundness of $\tm$ under the stated assumptions \citep{traceable-mixnets}, sets $\C_{\text{FM}}$ and $\V_{\text{FM}}$ correctly represent the set of true failures. This, combined with the definitions of $\R_{\text{FI}}$, $\C_{\text{FI}}$, $\V_{\text{FO}}$ and $\Pa_{\text{FO}}$, implies that the number of real failures in $(R^*,C^*,V^*,P^*)$ is less than the winning margin.
 	
	In the second case, $J^*$ is, as required, exactly the non-empty set of booths that contributed some failing item in $R_{\text{F}}$, $C_{\text{F}}$ (by the definitions on line 12), $P_{\text{F}}$ (by line 13 and the computational binding of Pedersen commitments under the DL assumption in $\Gone$) or $V_{\text{FO}}$ (by lines 14-18 and the soundness property of $\tm$; note that $V_{\text{FM}}$ entries in $V_{\text{F}}$ are mix-server errors and, as required, are not reported here). Finally, by line 21 and the soundness of $\tm$, $V^*$ is exactly the set of votes decrypted from encrypted votes sent by booths outside $J^*$. Thus, by the definitions of $R^*$, $C^*$ and $P^*$, $(R^*,C^*,V^*,P^*)$ does not contain any failures.
\end{proof}

\begin{theorem}
	Under the DDH assumption in $\Gone$ and the DCR assumption \citep{paillier}, the \emph{OpenVoting} protocol satisfies vote secrecy with recoverability as per Definition \ref{def:secrecy-mod-recoverability} in the random oracle model. 
\end{theorem}
\begin{proof}[Proof (Sketch)]
    If the adversary corrupts $\gen{2}$ but not $\gen{1}$, then it does not learn the ballot secrets of ballots used by $\voter{i_0}$ and $\voter{i_1}$ by the perfect hiding of Pedersen commitments under the DDH assumption and the IND-CPA security of Paillier schemes $\pke$ and $\thez$ under the DCR assumption (see Figure \ref{fig:ballotgen-protocol}). Post-printing, ballots get sealed and are opened only by the voter during vote casting, where the adversary-controlled BMD does not see any information about the ballot used. The receipts and the tallying protocol does not reveal any information to the corrupted mix-servers by the secrecy property of $\tm$ under the stated assumptions \citep{traceable-mixnets}. VVPRs are collected after mixing and the ballot identifiers used therein cannot be linked to the identifiers of $\voter{i_0}$ and $\voter{i_1}$. Finally, during the $\audit$ protocol, it is required that if the adversary causes a failure in either the receipt, encrypted vote or VVPR contributed by $\voter{i_0}$'s booth $j_0$ then it should also cause a failure in $\voter{i_1}$'s booth $j_1$. Thus, sets $R_{\text{FI}}$ to $P_{\FO}$ in Figure \ref{fig:audit-protocol} do not help it distinguish between the two worlds. Outputs $V_{\text{F}_j}$ do not help because for each failure in booth $j_0$, the adversary is required to create an identical failure in booth $j_1$. Further, the partial tally $V^*$ includes either both $v_0,v_1$ or none of them. The secrecy property of $\tm$ ensures that no additional information beyond the query outputs is revealed.

	If the adversary corrupts $\gen{1}$ but not $\gen{2}$, then it obtains ballot secrets but it cannot identify which of $\voter{i_0}$ or $\voter{i_1}$ used which ballot. The rest of the proof is similar.
\end{proof}

\section{Conclusion and Future Work}

We have introduced and formalised the notion of recoverability and secrecy for dual voting protocols and suggested a protocol that achieves this notion. Based on existing reports for the underlying traceable mixnet construction, the total time taken by the recovery process remains within a few hours for $n=10000$ ciphertexts, which can be optimised further using the construction's high degree of task parallelism \citep{traceable-mixnets}. 

Although we have shown our protocol's recoverability properties, the potential non-termination of the revoting process during vote casting seems like an inherent limitation of BMD protocols and designing voting frontends that overcome this limitation yet remain usable and minimise voter errors appears to be a challenging open problem. Further, although we have focused on recoverability for first-past-the-post voting where exact winning margins are computable, extending to other more complex voting rules also appears to be an interesting avenue for future work. \\

\noindent \textbf{Acknowledgments.} Prashant Agrawal is supported by the Pankaj Jalote Doctoral Grant. Abhinav Nakarmi was supported by a research grant from MPhasis F1 Foundation.

\bibliography{recoverability}
\bibliographystyle{plain}

\end{document}